\documentclass[prb,aps,amsmath,preprint]{revtex4}

\usepackage{graphicx}

\usepackage{dcolumn}

\usepackage{bm}

\usepackage{color}

\begin{document}

\title{Elastic constants and high-pressure structural transitions in lanthanum monochalcogenides from
experiment and theory}

\author{G. Vaitheeswaran$^{1,*}$, V. Kanchana$^{1}$, S. Heathman$^{2}$, 
M. Idiri$^{2}$, T. Le Bihan$^{3}$, A. Svane$^{4}$, A. Delin$^{1}$ and B. Johansson$^{1,5}$} 
\affiliation{$^{1}$Applied Materials Physics, Department of Materials Science and Engineering, 
Royal Institute of Technology, Brinellv\"agen 23, 100 44
Stockholm, Sweden \\
$^{2}$European Commission, JRC, Institute for Transuranium Elements, Postfach 2340, D-76125, Karlsruhe, Germany \\
$^{3}$European Synchrotron Radiation Facility, BP220, F-38043 Grenoble Cedex, France \\
$^{4}$ Department of Physics and Astronomy, University of Aarhus, DK-8000 Aarhus C, Denmark\\
$^{5}$Condensed Matter Theory Group, Department of Physics, Uppsala University, Box.530, SE-751 21, Uppsala, Sweden}
\date{\today}

\begin{abstract}
The high-pressure structural behavior of lanthanum monochalcogenides is investigated by theory and experiment.
Theory comprises density functional calculations of LaS, LaSe and LaTe with the general gradient approximation 
for exchange and correlation effects, as implemented within the full-potential linear muffin-tin orbital method. 
The experimental studies consist of high-pressure angle dispersive x-ray 
diffraction investigations of LaS and LaSe up to a maximum pressure of 41 GPa. 
A structural phase transition from the NaCl type to CsCl type crystal structure is found to occur in all cases. 
The experimental transition pressures are 27-28 GPa and 19 GPa, for LaS and LaSe, respectively, 
while the calculated transition pressures are 29 GPa, 21 GPa and 10 GPa for LaS, LaSe and LaTe, respectively.
The calculated ground state properties such as equilibrium lattice constant, bulk modulus and its 
pressure derivative, and Debye temperatures are in good agreement with experimental results. 
Elastic constants are predicted from the calculations.
\end{abstract}

\pacs{61.50.Ks, 71.15.Nc, 62.20.Dc, 71.20.Gj}
\maketitle

\section {Introduction}
The lanthanum monochalcogenides belong to the wide class of binary rare-earth monochalcogenides with the 
NaCl-type structure, which has been intensively studied because of their interesting physical properties 
including complex magnetic structures, pressure induced insulator-metal transitions, 
anomalous valence fluctuations, and unusual Fermi surface properties.\cite{Dernier,Benedict} 
The trivalent lanthanum monochalcogenides are superconductors near 1 K, and the superconducting 
transition temperature as well as the electronic specific heat coefficient increases from the 
monosulfide to the monotelluride, whereas the Debye temperature decreases from LaS to LaTe\cite{Bucher}. 

In this work we explore the high pressure behavior of the La chalcogenides by experiment and theory. 
High-pressure x-ray diffraction experiments are conducted on LaS and LaSe to obtain equations of states 
including discontinuous structural transitions from the NaCl structure 
(Space group Fm3m, Z=1, also called the B1 structure in the following) 
to the CsCl structure 
(Space group Pm3m, Z=4, also termed the B2 structure in the following).
Density functional calculations are performed to compare basic theoretical predictions with the measurements. 
The present experimental study of LaS extends our previous study on U$_{x}$La$_{1-x}$S compounds\cite{Bihan}. 
For LaSe no high-pressure results have been reported previously, while high-pressure experiments on 
LaTe report a structural transition from B1 to B2 at around 7 GPa\cite{Jayaraman}.

A second objective of the present work is to investigate the elastic properties of the 
lanthanum chalcogenides, for which there are no experimental results available. 
Elastic constants are derived from total energy variations with applied strains. 
We compare our results with recent theoretical work\cite{theochem} and find good agreement.


The systematics of the electronic band structure through the lanthanum and cerium chalcogenide 
series was studied with angle-resolved photoemission spectroscopy\cite{Nakayama,Nakayama2}. 
The Fermi surface properties of lanthanum monochalcogenides have been studied by de Haas-van Alphen 
effect measurements\cite{Kimura,Kimura2}. From the theoretical side the energy band structure, 
superconductivity, surface electronic structure, optical and magneto-optical spectra as well as 
structural stability of lanthanum monochalcogenides have been investigated by several authors 
using the local spin density approximation (LSDA) as well as the LSDA+U 
approximation\cite{Vlasov,Lu,Sankaralingam,Olle1,Olle2,Vaithee,Antonov}. 
A few papers also focused on the calculation of second and third order elastic constants 
using the short range repulsive potential method \cite{Yadav,Dinesh,Dinesh1}. 
A few experimental studies such as point contact spectroscopy\cite{Wachter}, reflectivity\cite{Kaldis} 
and phonon spectra\cite{Steiner} have been reported for LaS. The magneto-optical properties of 
lanthanum chalcogenides have been investigated experimentally,\cite{Pittini1,Pittini2,Wachter2} 
where a non-zero Kerr effect could be observed in these Pauli paramagnets by applying an external magnetic field.

The rest of the paper is organized as follows. In section 2 we discuss the computational and 
experimental details of our work. The calculated ground state properties, elastic constants as well as 
the experimental and theoretical results for the high-pressure behavior and structural transitions
are presented in Section 3, and the conclusions are given in Section 4. 

\section {Computational and Experimental details}
\subsection{The electronic structure method}
In this work we have used the all-electron full-potential linear muffin tin orbital (FP-LMTO) 
method\cite{Savrasov} to calculate the total energies and the basic ground state properties. 
Here the crystal is divided into two regions: non-overlapping muffin-tin spheres surrounding 
each atom and the interstitial region between the spheres. We used a double $\kappa$ spdf LMTO 
basis (each radial function within the spheres is matched to a Hankel function in the 
interstitial region) for describing the valence bands. The following basis set was used in the
calculations: La(5s,6s,5p,5d,4f), S(3s,3p,3d), Se(4s,4p,3d,4d), Te(5s,5p,4d,5d).
The exchange correlation potential was calculated within the generalized gradient approximation 
(GGA) scheme\cite{Perdew}. The charge density and potential inside the muffin-tin spheres 
were expanded in terms of spherical harmonics up to $l_{max}$=6, 
while in the interstitial region, they were expanded in plane waves, with 
6566 (energy up to 109.80 Ry) waves being included in the calculation. 
Total energies were calculated as a function of volume, for a (18 18 18) k-mesh,
corresponding to 195 k-vectors in the irreducible wedge of the Brillouin zone,
 and the results fitted to the Birch equation of state\cite{Birch} to obtain the ground state properties. 

The elastic constants were obtained from the variation of the total energy under volume-conserving strains, 
as outlined in Ref. \onlinecite{oxides}.

\subsection{Experimental details}
Experiments were carried out at the European Synchrotron Radiation Facility (ESRF), 
on the ID30 beamline dedicated to high pressure diffraction experiments, in the angular 
dispersive mode, using a double focused monochromatic beam at $\lambda$=0.3738 \AA. Loading of 
LaS and LaSe powders were performed in Le Toullec-type diamond anvil cells (DAC), using nitrogen, 
argon or silicone oil as the pressure transmitting media. The pressure inside the cell was determined 
via the ruby fluorescence method \cite{Mao} from a small ruby ball mounted alongside the LaX sample. 
The sample to detector distance was calibrated before each 
set of experiments by means of a standard silicon powder sample.
Diffraction images were captured with a FASTSCAN \cite{Thoms} image plate detector, and processed 
using the ESRF Fit2D program\cite{Hanfland} to provide data in a format suitable for 
Rietveld analysis using the FULLPROF program\cite{Rodriguez}. 
The lattice constants 
at ambient pressure have been determined as a$_0$=5.852 and a$_0$=6.067 \AA\ for LaS and LaSe, 
respectively, in good agreement with previous experimental data given by Ref. \onlinecite{Vaithee}.

\section{Structural, Elastic and High pressure studies} 
\subsection{Ground state and Elastic properties}
The calculated ground state properties such as equilibrium lattice constant, bulk modulus 
and its pressure derivative are given in Table 1. The calculated equilibrium lattice constants
are generally overestimated  by $\sim 0.5$ \% compared to experiment, which is an improvement
compared to earlier calculations\cite{Lu,Vaithee}. Similarly, for the bulk modulus, the 
agreement between theory and experiment has improved\cite{Lu,Vaithee}. 
It is interesting to note that the bulk moduli of the lanthanum chalcogenides are quite similar to
those of the neighboring cerium chalcogenides, which are characterized by a localized $f$ state\cite{Leger}: 
$B= 82$ GPa, $76$ GPa and $58$ GPa, for CeS, CeSe and CeTe, respectively. 
Similarly, the calculated bulk moduli of the praseodymium chalcogenides are $B= 89$ GPa, $78$ GPa and $57$
GPa, for PrS, PrSe and PrTe, respectively, when Pr is represented in a trivalent configuration \cite{PrX}.
These similarities of numbers corroborate 
the practice of taking the lanthanum chalcogenides as non-magnetic reference systems for
the later rare-earth chalcogenides.


The calculated elastic constants of LaS, LaSe and LaTe are listed in Table 2, where  we also 
compare to the recent LAPW calculation of Ref. \onlinecite{theochem}. 
The two theoretical approaches find very similar $C_{11}$ value for LaS and LaSe, while for LaTe 
this parameter is $\sim 13 \%$ smaller in the present calculations than found by the LAPW method. The 
present $C_{12}$ and $C_{44}$ parameters are in all cases lower than found in Ref. \onlinecite{theochem}, for   
LaSe    $C_{12}$ with  almost a factor of 2 difference.
Our calculated shear moduli $G$ are 19\% to 40\% larger than the ones calculated with LAPW.
For Young's modulus $E$, our values are 16\% to 35\% higher, and finally, for Poisson's ratio $\nu$ our
values are 8\% to 17\% lower than the LAPW results.
The largest discrepancies between the two calculations occur
for LaTe in all cases.
Furthermore, we observe the general trend that the heavier the chalcogenide is, the softer is the compound.
This trend is expected and easy to understand, since with increased chalcogen size,
the lattice parameter increases and the valence orbitals become increasingly delocalized due to a higher
number of nodes.\cite{AD}


As previously stated, there is at present no experimental information regarding the elastic constants 
available for the lanthanum chalcogenides.
The elastic constants of LaS are of roughly the same 
magnitude as those measured for the isoelectronic YS compound (Table 2). 

Table 3 presents sound velocities and Debye temperatures, as derived from the calculated elastic 
constants\cite{oxides}. The Debye temperatures have been determined experimentally from the
low-temperature specific heat, and the agreement with our calculations is excellent
(the calculated Debye temperatures are 0.5\% to 6\% higher than experiment),
 which can be taken as an indirect check on our calculated elastic constant values.

\subsection{High-pressure structural transitions}
\begin{figure}
\includegraphics[width=\linewidth,clip, width=9.0 cm]{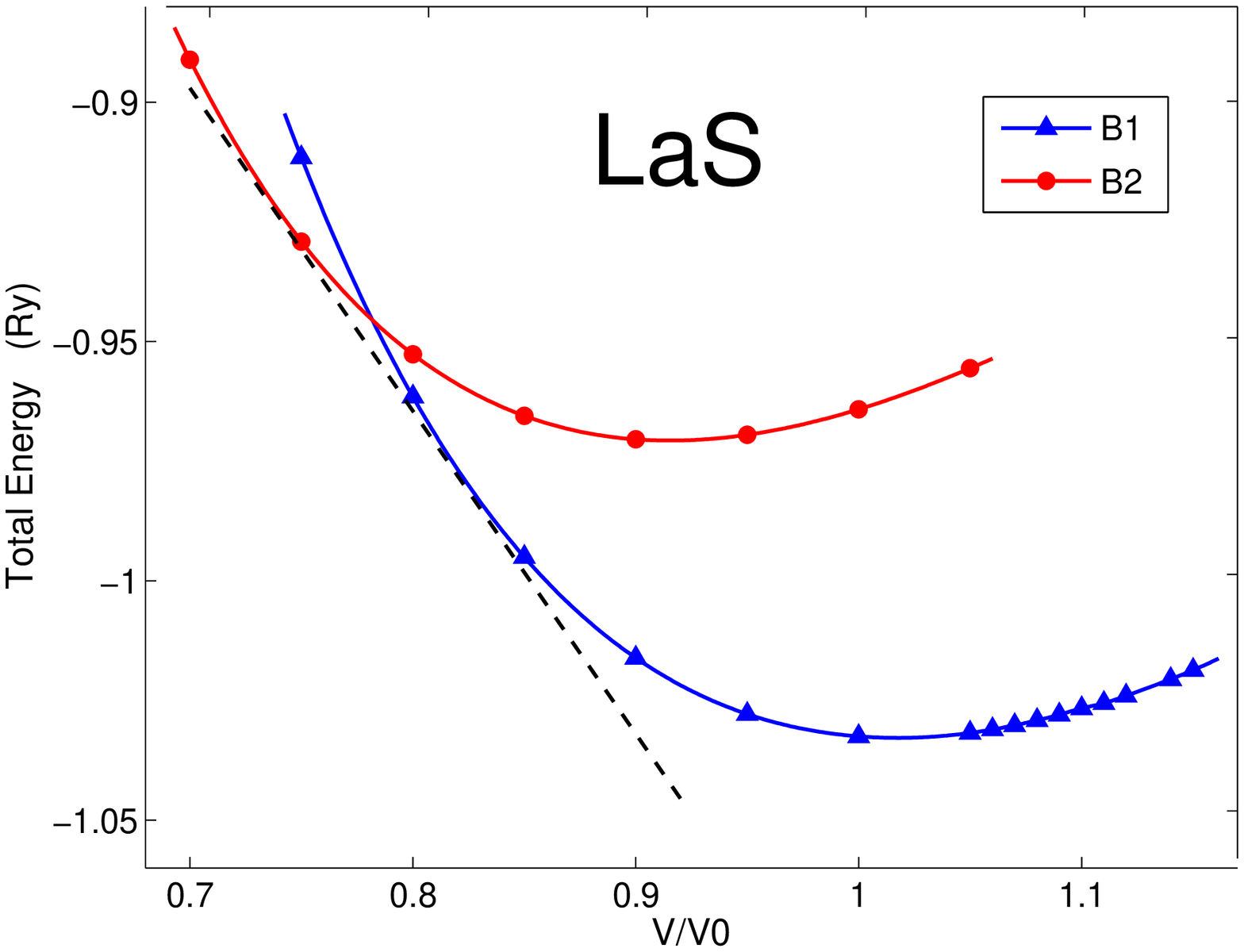}
\includegraphics[width=\linewidth,clip, width=9.0 cm]{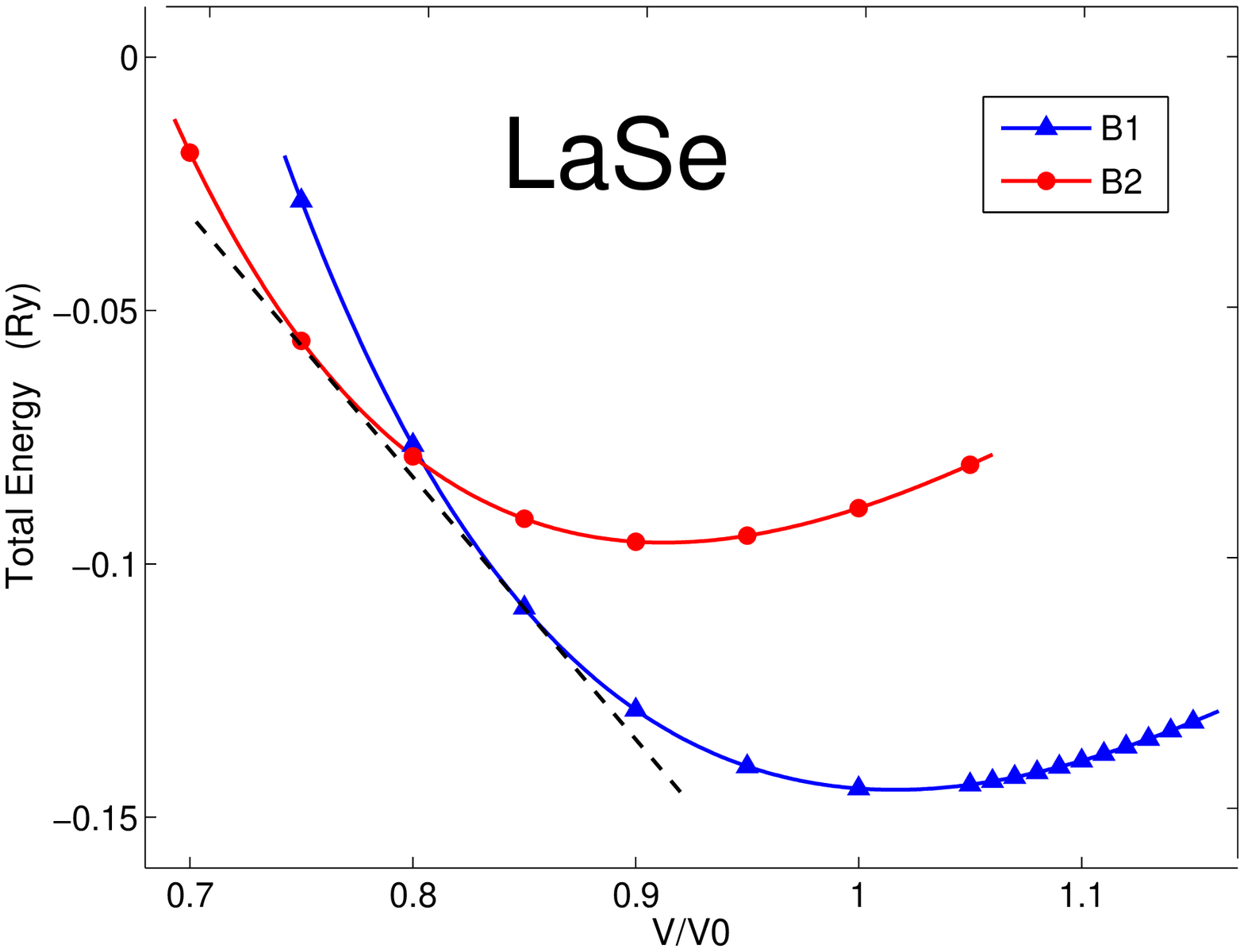}
\includegraphics[width=\linewidth,clip, width=9.0 cm]{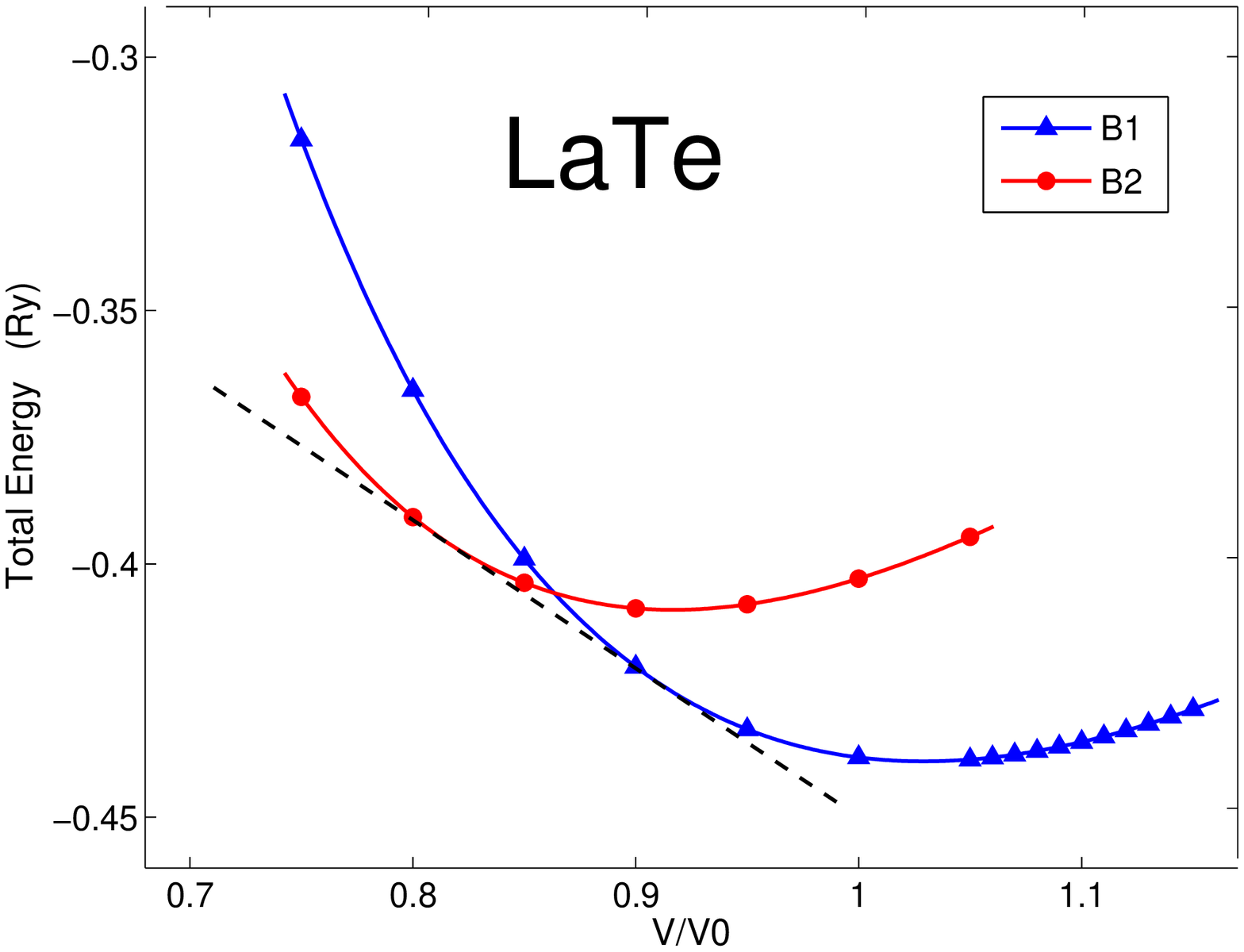}
\caption{(Color online) Calculated total energy vs. relative volume for (a): LaS, (b): LaSe, and (c): LaTe
 in the B1 and B2 structures. The common tangents mark the structural B1 $\rightarrow$ B2 
transition.}
\label{fig:1}
\end{figure}

\begin{figure}
\includegraphics[width=\linewidth,clip, width=10.0 cm]{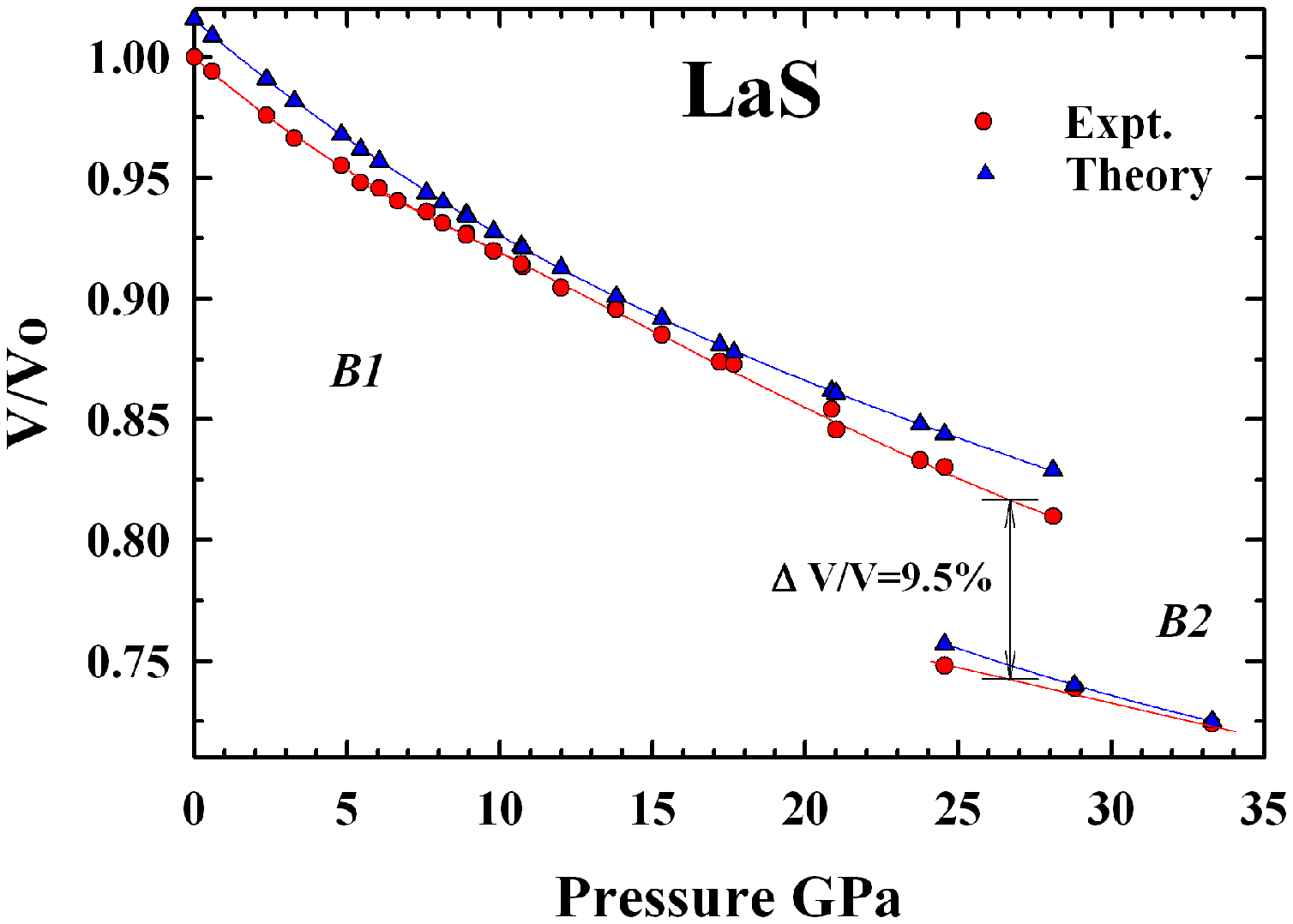}
\caption{(Color online) Pressure as a function of 
relative volume for LaS in the B1 and B2 structures. Triangles:
theory (GGA), circles: experiment. The experimental volume collapse at the B1 $\rightarrow$ B2 transition
is marked.}
\label{fig:2}
\end{figure}

The high pressure structural behavior of the lanthanum chalcogenides was studied both experimentally and 
theoretically. The calculated total energies as functions of relative volume for 
LaS, LaSe and LaTe are shown in figure \ref{fig:1}, while experimental and theoretical pressure-volume
relations are presented in figures 2, 4 and 5 for LaS, LaSe and LaTe, respectively. Figure 3 shows the
diffraction spectra recorded for LaS and LaSe.
It appears from the theory presented in figure 1 
that with compression the B2   phase 
 becomes more and more favorable, and eventually
a transition from the B1   structure to the B2   structure occurs. 
From the common tangent of the B1 and B2 total energies the transition pressure
is determined.
The calculated transition pressure for LaS is 29.3 GPa with a volume collapse of 10.3\%,
 which is in good agreement with the experimental transition pressure of 27-28 GPa 
with a volume collapse of around 9.5\% as shown in figure \ref{fig:2}.
A similar transition was predicted to occur in CeS around 24.3 GPa,
 in which the Ce ions are in a tetravalent state\cite{Axel} In the case of PrS a similar
transition is predicted to occur around 22 GPa, in this case with the Pr ions remaining
 trivalent\cite{PrX}. The transition pressure of LaS is similar  to  that of the pnictogen group neighbour 
LaP, in which  a transition from the B1 structure to a distorted B2 structure is observed 
experimentally around 24 GPa\cite{Adachi}. 

Like LaS, LaSe also undergoes a B1 $\rightarrow$ B2 transition, experimentally seen at a pressure of
19 GPa with a volume reduction of 10\%, cf. figure \ref{fig:4}, for which the present theory finds concordant values of
$P_t = 21$ GPa with a volume collapse of $10.4\%$. 
In the case of CeSe a similar transition from B1 type to B2 type occurs around 20 GPa\cite{Leger}, in
this case with the Ce ion remaining trivalent on both sides of the transition according to theory\cite{Axel}. 
Even for PrSe a B1 $\rightarrow$ B2 transition is predicted to occur around 12 GPa, wherein the Pr ions remain trivalent\cite{PrX}. 
The transition pressure of LaSe is similar to that of the pnictogen neighbour LaAs, in which a transition from B1 structure to 
a distorted B2 type occurs around 20 GPa\cite{shirotani}.

As far as LaTe is concerned a transition from  B1  type to  B2  type is reported around 7 GPa\cite{Jayaraman},
which agrees quite well with the calculated transition pressure of 9.7 GPa with a volume reduction of 10.4\%, cf. figure 5. 
Both CeTe and PrTe undergo a transition from B1 type to B2 around 8$\pm$1 GPa\cite{leger2} and 9$\pm$1 GPa\cite{prte}. When 
comparing the pnictogen neighbour LaSb with that of LaTe the pnitogen undergoes a transition from B1 type to a distorted B2 type 
structure around 11 GPa\cite{hayashi} which is slightly higher than that of LaTe. 

\section{Conclusions}

By means of a combined theoretical and experimental study the high-pressure structural behavior of LaX 
(X=S, Se, Te) compounds have been investigated. Unlike the lanthanum monopnictides which show a transition from B1 to
  a distorted B2 structure\cite{LaP,LaSb}, the lanthanum monochalcogenides exhibit a simple B1 $\rightarrow$ 
B2 structural phase transition,  similar to the one found for most of the lanthanide and actinide 
monochalcogenides studied up to now\cite{Benedict,Steve,Gensini,Leif,Le,Le2}.

For all three systems studied here, the volume collapse is around 10\% (from both experiment and theory)
at the B1 $\rightarrow$ B2 transition.
Let us compare this magnitude of the volume collapse to a very simple model in which we assume
that at the transition point, the La--X
bond length is restored to the ambient value. We expect such a model to underestimate the volume collapse,
since the
 bond
length at high pressure most probably will be a bit smaller than at ambient pressure.
Our simple model gives the following relation between the
relative volume $V_t^{B1}/V_0^{B1}$ just before the transition and the volume collapse
$1-V_t^{B2}/V_t^{B1}$:
\begin{equation}
1-\frac{V_t^{B2}}{V_t^{B1}} = 1-\frac{4}{3\sqrt{3}} \frac{V_0^{B1}}{V_t^{B1}}.
\end{equation}
For LaS, the experimental relative volume at the transition is 0.82, which would give a volume collapse of
6\% according to this model. For LaSe, the corresponding numbers are
0.84 and 8\%. Thus, the volume collapse observed experimentally (and from density functional calculations)
can be viewed as a partial restoration
of the original bond length between the La and the chalcogenide.
Based on the above discussion we conclude that the volume collapse observed in the lanthanum
chalcogenides is entirely consistent with a simple picture of the transition in which the
volume collapse is a consequence of the rearrangement of the atoms
to a more close-packed structure while the total number of valence electrons remains unchanged.
We have also checked the partial occupation number of spdf states across the structural transition and there is no 
appreciable change in the occupation numbers across the transition.
This is in contrast to the situation in many Ce systems,
in which the volume collapse is accompanied by delocalization
of an $f$ electron.\cite{borjephilmag,Delin97,Axel,cep}
 
\acknowledgments

G. V,  V. K, A. D, and B. J  acknowledge V. R. and SSF for the financial support and SNIC for the computer time. The authors
S.H. , M.I. and T.L.B. would like to thank K. Mattenberger and O. Vogt (ETH, Z\"urich) for the samples used in this 
study. Support given to M.I. within the framework of the EC funded program "Human capital and mobility" is also acknowledged.

\newpage

\begin{table}[tb]
\caption{
Calculated lattice constants in \AA, bulk modulus in GPa,
and its pressure derivative $B' $
of the lanthanum monochalcogenides in the B1   structure}
\begin{ruledtabular}
\begin{tabular}{cccccc}
Compound   &      &Lattice Constant  & Bulk Modulus     & $B'$   \\
&&&&&\\
\hline

LaS  & Expt. this work   & 5.852                                     & 89(3),83.6$^d$              & 6.5(4) \\
     & Theory, this work & 5.873                                     & 87.8                        &  3.95\\
     & Other theory      & 5.812$^a$,5.773$^b$,5.727$^c$,5.895$^e$   & 97.74$^a$,107$^c$,81.53$^e$ &  4.67$^e$   \\

LaSe & Expt. this work   & 6.067                          &74(2)               & 4.7(3) \\
     & Theory, this work & 6.091                          &74.8                & 4.12  \\
     & Other theory      & 5.957$^c$,6.126$^e$           &97.74$^c$,68.40$^e$ & 4.28$^e$ \\ 
     
LaTe & Expt.             & 6.435$^h$                      & 60.6$\pm$2$^f$,55$^g$   & - \\
     & Theory, this work & 6.470                          & 59.4             & 4.12 \\
     & Other theory      & 6.255$^c$,6.512$^e$            & 74.02$^c$,55.34$^e$    & 4.96$^e$ \\              
\end{tabular}
\end{ruledtabular}
$^a$Ref.\onlinecite{Lu}; $^b$Ref.\onlinecite{Olle1}; $^c$Ref.\onlinecite{Vaithee}; 
$^d$Ref.\onlinecite{Holtzberg}; $^e$Ref.\onlinecite{theochem}; $^f$Ref.\onlinecite{Jayaraman2}; 
$^g$Ref. \onlinecite{Benedict};
$^h$Ref. \onlinecite{Kimura2}.
\end{table}

\begin{table}[tb]
\caption{
Calculated elastic constants, shear modulus $G$, Young's modulus $E$  - all given in GPa  - 
and Poisson's ratio $\nu$
for lanthanum monochalcogenides in the B1   structure at the theoretical equilibrium volume.
For comparison, results of LAPW calculations (Ref. \onlinecite{theochem}) 
and experimental results for YS (Ref. \onlinecite{Weber}) are included.
}
\begin{ruledtabular}
\begin{tabular}{cccccccc}
 Compound     &$C_{11}$  &$C_{12}$  &$C_{44}$ &  G    &E      &$\nu$  &  \\ \hline

LaS           & 227.9    & 18.0     &  22.2   & 55.3  & 137.2 & 0.240 & Present   \\
              & 234      & 23       &  25     & 46.6  & 117.8 & 0.26  & LAPW      \\
YS            & 250      & 20       &  30     &  -    &  -    &  -    & Expt.     \\
LaSe          & 201.6    & 11.4     &  15.7   & 47.5  & 117.5 & 0.238 & Present   \\
              & 203      & 21       &  22     & 40.6  & 102.5 & 0.26  & LAPW      \\
LaTe          & 158.7    &  9.7     &   7.9   & 34.5  &  86.8 & 0.256 & Present   \\
              & 171      & 12       &   8     & 24.6  &  64.1 & 0.30  & LAPW      \\
\end{tabular}
\end{ruledtabular}
\end{table}

\begin{table}[tb]
\caption{
Calculated longitudinal, shear, and average wave velocity ($v_l$, $v_s$, and $v_m$, respectively) 
in m/s, and the Debye temperature, $\theta_D$, in kelvin from the average elastic wave velocity 
for lanthanum monochalcogenides in the B1 structure at the theoretical equilibrium volume.}
\begin{ruledtabular}
\begin{tabular}{ccccccc}
 Compound    &     &\it{v$_l$}  &\it{v$_s$}  &\it{v$_m$} & $\theta_D$   \\ \hline

LaS           & Present & 5249     & 3072 &3406 & 277.2      \\
              & Expt.$^a$   &  -         &  -     & -  & 276     \\  
LaSe          & Present &4626      & 2712 &3006 & 233.6  \\
              & Expt.$^a$   &  -         &   -   &  -    & 231 \\  
LaTe          & Present & 3972     & 2273 &2524 & 185.7      \\
              & Expt.$^a$   &  -          & -    & -   & 175 

\end{tabular}
\end{ruledtabular}
$^a$Ref.\onlinecite{Bucher}
\end{table}

\begin{table}[tb]
\caption{Calculated and experimental transition pressures, $P_t$ in GPa, and volume changes,
given in \%, for the B1 $\rightarrow$ B2 structural phase transition of lanthanum monochalcogenides.}
\begin{ruledtabular}
\begin{tabular} {ccccc}
Compound & \multicolumn{2}{c}{$P_t$} & \multicolumn{2}{c}{Volume Collapse}  \\ 
         & Theory & Expt.            & Theory & Expt. \\
\hline
LaS   & 29.3  & 27-28     & 10.3  &  9.5  \\
LaSe  & 21    & 19        & 10.4  &  10.5 \\
LaTe  & 9.7   &  7$^a$     & 10.4 &   -    \\ 

\end{tabular}
\end{ruledtabular}
$^a$Ref.\onlinecite{Jayaraman}
\end{table}

\begin{figure}
\includegraphics[width=70mm,clip, width=9.0 cm]{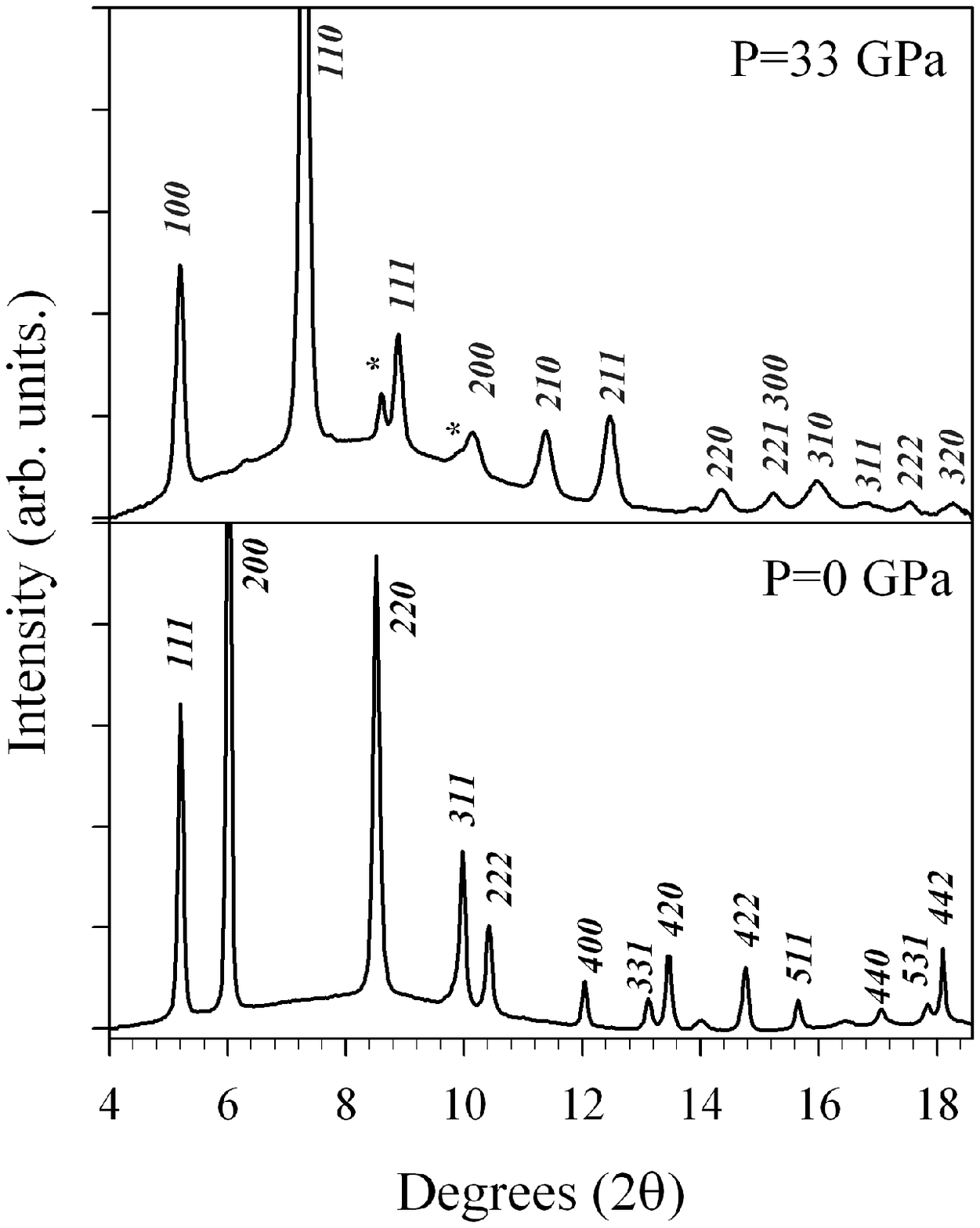}
\includegraphics[width=70mm,clip, width=9.0 cm]{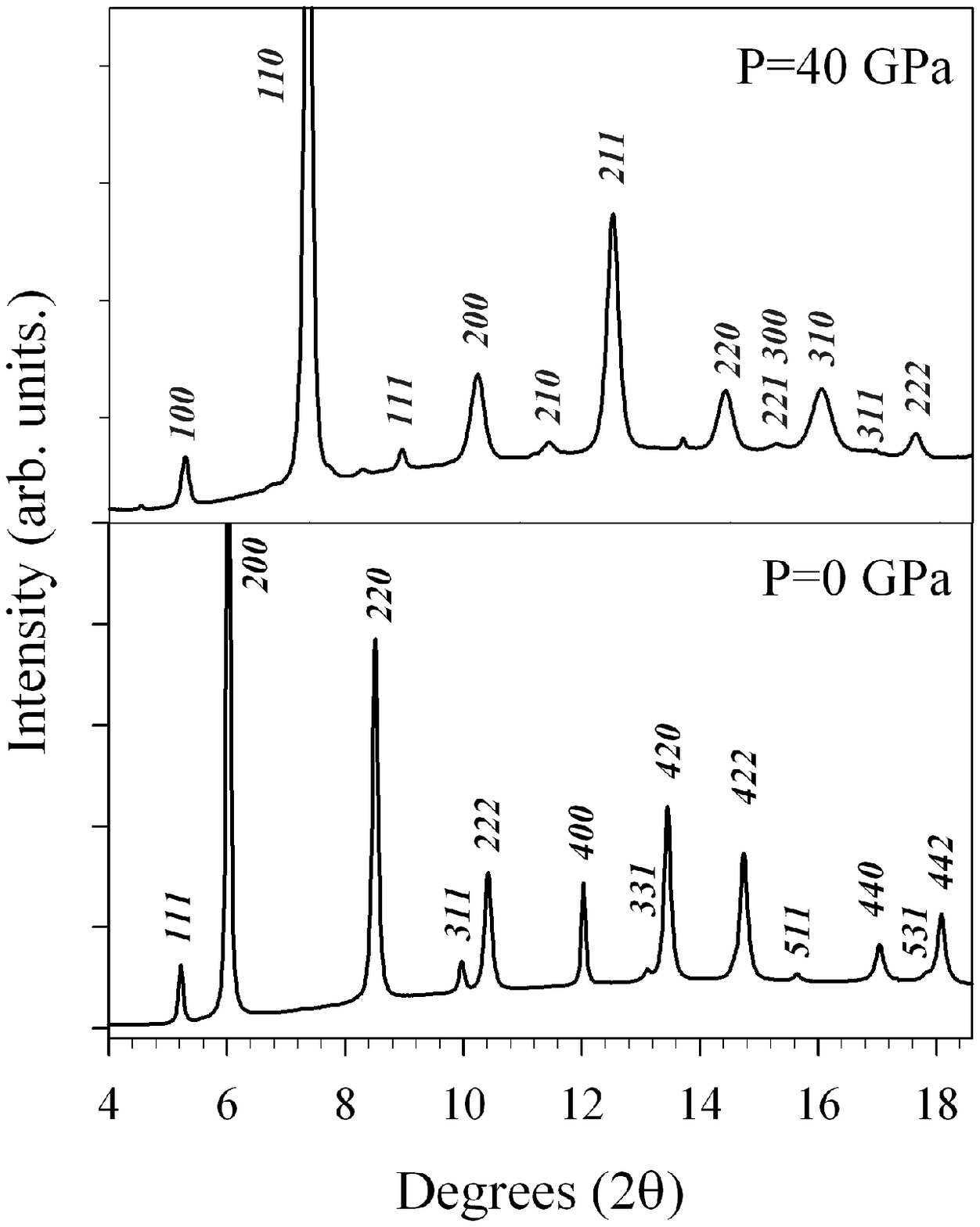}
\caption{X-ray diffraction spectra of (a): LaS and (b): LaSe in the B1 structure at 
$P=0$ GPa, $a=5.8520$ \AA\ for LaS, $a=6.0670$ \AA\ for LaSe, and in the B2 structure right after the
transition,
at $P=33$ GPa and  $a=3.3068$ \AA\ for LaS, and at $P=40$ GPa and $a=3.3684$ \AA\ for LaSe.}
\label{fig:3}
\end{figure}

\begin{figure}
\includegraphics[width=\linewidth,clip, width=10.0 cm]{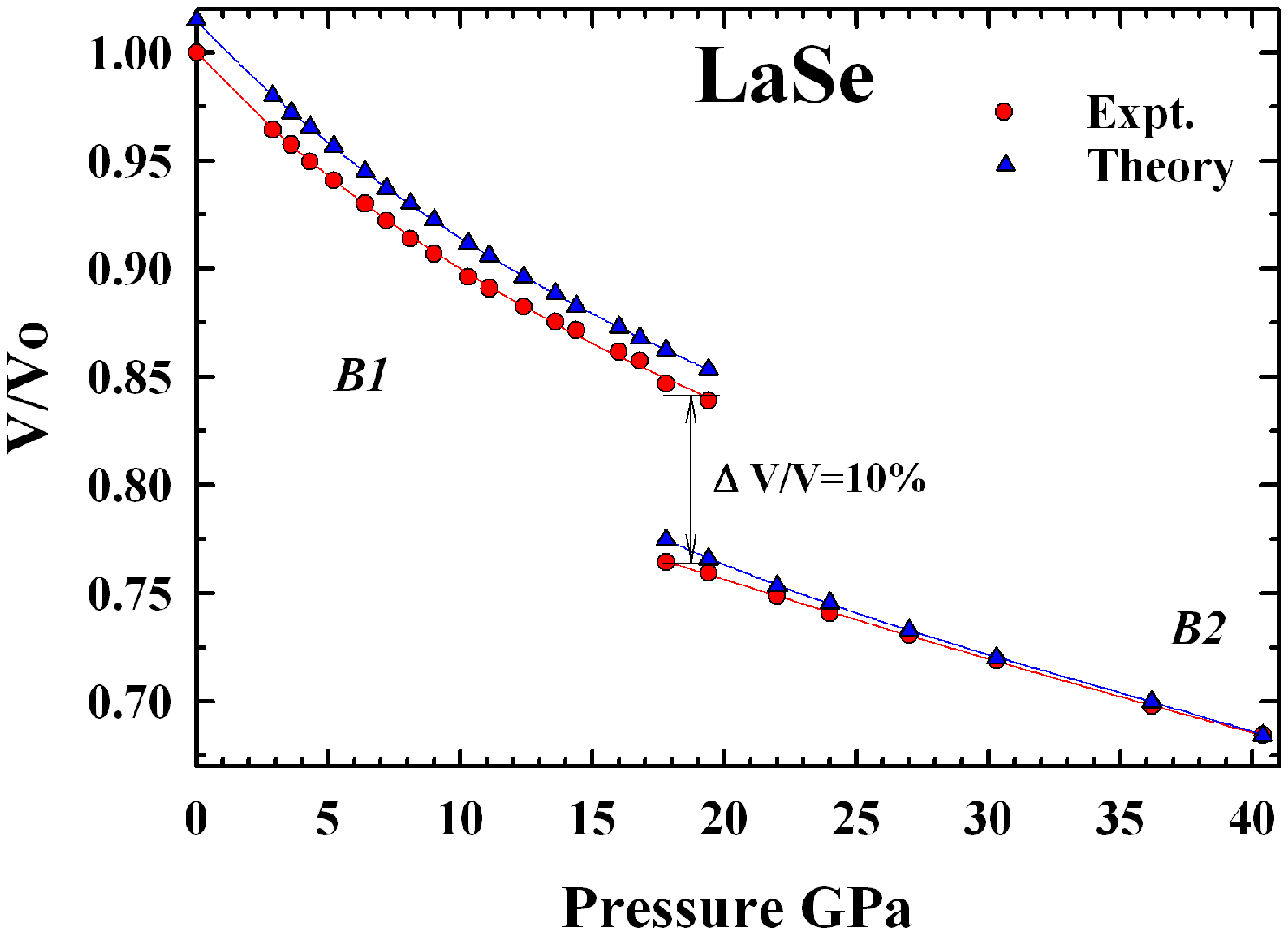}
\caption{(Color online) Pressure as a function of relative volume for LaSe in the B1 and B2 structures.
Triangles:
theory (GGA), circles: experiment. The experimental volume collapse at the B1 $\rightarrow$ B2 transition
is marked.}
\label{fig:4}
\end{figure}

\begin{figure}
\includegraphics[width=\linewidth,clip, width=10.0 cm]{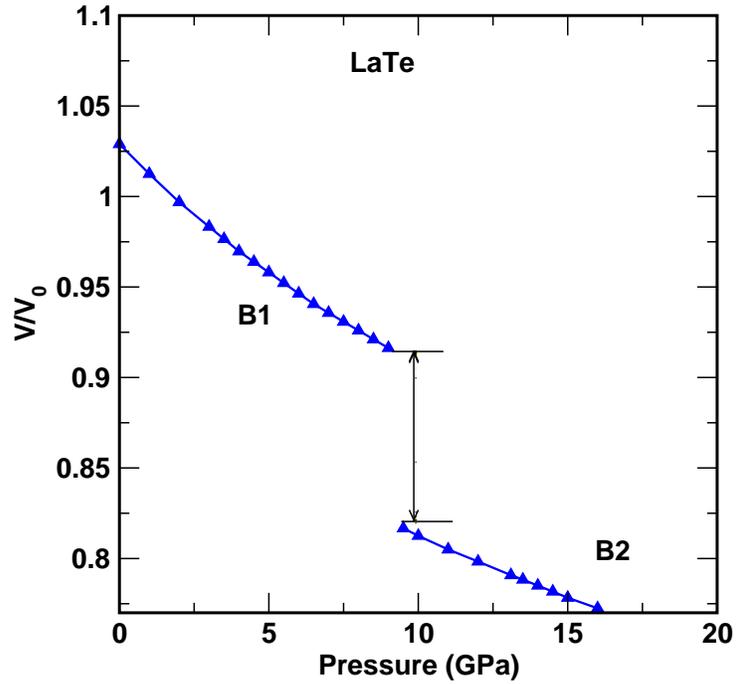}
\caption{(Color online) Theoretical pressure as a function of relative volume for LaTe in the B1 and B2 structures.}
\label{fig:5}
\end{figure}

\end{document}